# A Survey on Optimization Approaches to Text Document Clustering

R.Jensi[1] and Dr.G.Wiselin Jiji[2]

[1]Research Scholar, Manomanium Sundaranar University, Tirunelveli, India
[2]Dr.Sivanthi Aditanar College of Engineering, Tiruchendur, India

### ABSTRACT

*Text Document Clustering is one of the fastest growing research areas because of availability of huge amount of information in an electronic form. There are several number of techniques launched for clustering documents in such a way that documents within a cluster have high intra-similarity and low inter-similarity to other clusters. Many document clustering algorithms provide localized search in effectively navigating, summarizing, and organizing information. A global optimal solution can be obtained by applying high-speed and high-quality optimization algorithms. The optimization technique performs a globalized search in the entire solution space. In this paper, a brief survey on optimization approaches to text document clustering is turned out.*

### KEYWORDS

*Text Clustering, Vector Space Model, Latent Semantic Indexing, Genetic Algorithm, Particle Swarm Optimization, Ant-Colony Optimization, Bees Algorithm.*

## 1. INTRODUCTION

With the advancement of technologies in World Wide Web, huge amounts of rich and dynamic information's are available. With web search engines, a user can quickly browse and locate the documents. Usually search engines returns many documents, a lot of which are relevant to the topic and some may contain irrelevant documents with poor quality. Cluster analysis or clustering plays an important role in organizing such massive amount of documents returned by search engines into meaningful clusters. A cluster is a collection of data objects that are similar to one another within the same cluster and are dissimilar to objects in other clusters. Document clustering is closely related to data clustering. Data clustering is under vigorous development. Cluster analysis has its roots in many data mining research areas, including data mining, information retrieval, pattern recognition, web search, statistics, biology and machine learning. Even if a lot of significant research effort has been done in [3,15,20,22,26,33], the more challenges in clustering is to improve the quality of clustering process.

In machine learning, **clustering** [3] is an example of **unsupervised learning**. In the context of machine learning, clustering contracts with **supervised learning** (or **classification).** Classification assigns data objects in a collection to target categories or classes. The main task of classification is to precisely predict the target class for each instance in the data. So classification algorithm requires training data. Unlike classification, clustering does not require training data. Clustering does not assign any per-defined label to each and every group. Clustering groups a set of objects and finds whether there is *some* relationship between the objects.

As stated by [3], [29], Data clustering algorithms can be broadly classified into following categories:

- Partitioning methods
- Hierarchical methods
- Density-based methods
- Grid-based methods
- Model-based methods
- Frequent pattern-based clustering
- Constraint-based clustering

With partitional clustering the algorithm creates a set of data non-overlapping subsets (clusters) such that each data object is in exactly one subset. These approaches require selecting a value for the desired number of clusters to be generated. A few popular heuristic clustering methods are k-means and a variant of k-means-bisecting k-means, k-medoids, PAM (Kaufman and Rousseeuw, 1987), CLARA (Kaufmann and Rousseeuw, 1990), CLARANS (Ng and Han, 1994) etc.

With hierarchical clustering the algorithm creates a nested set of clusters that are organized as a tree. Such hierarchical algorithms can be agglomerative or divisive. Agglomerative algorithms, also called the bottom-up algorithms, initially treat each object as a separate cluster and successively merge the couple of clusters that are close to one another to create new clusters until all of the clusters are merged into one. Divisive algorithms, also called the top-down algorithms, proceed with all of the objects in the same cluster and in each successive iteration a cluster is split up using a flat clustering algorithm recursively until each object is in its own singleton cluster. The popular hierarchical methods are BIRCH [38], ROCK [39], Chamelon [37] and UPGMA.

An experimental study of hierarchical and partitional clustering algorithms was done by [37] and proved that bisecting kmeans technique works better than the standard kmeans approach and the hierarchical approaches.

Density-based clustering methods group the data objects with arbitrary shapes. Clustering is done according to a density (number of objects), (i.e.) density-based connectivity. The popular density-based methods are DBSCAN and its extension, OPTICS and DENCLUE [3].

Grid-based clustering methods use multiresolution grid structure to cluster the data objects. The benefit of this method is its speed in processing time. Some examples include STING, WaveCluster.

Model-based methods use a model for each cluster and determine the fit of the data to the given model. It is also used to automatically determine the number of clusters. Expectation-Maximization, COBWEB and SOM (Self-Organizing Map) are typical examples of model-based methods.

Frequent pattern-based clustering uses patterns which are extracted from subsets of dimensions, to group the data objects. An example of this method is pCluster.

Constraint-based clustering methods perform clustering based on the user-specified or application-specific constraints. It imposes user's constraints on clustering such as user's requirement or explains properties of the required clustering results.

A soft clustering algorithm such as fuzzy c-means [26], [42] has been applied in [30], [31], [43] for high-dimensional text document clustering. It allows the data object to belong to two or more clusters with different membership. Fuzzy c-means is based on minimizing the dissimilarity function. In each iteration it finds the cluster centroid that minimizes the objective function (dissimilarity function) [20].

## 2. SOFT COMPUTING TECHNIQUES

Soft computing [40] is foundation of conceptual intelligence in machines. Soft Computing is an approach for constructing systems which are computationally intelligent, possess human like expertise in particular domain, adapt to the changing environment and can learn to do better and can explain their decisions.Unlike hard computing, soft computing is tolerant of imprecision, uncertainty, partial truth, and approximation. In effect, the role model for soft computing is the human mind.

The realm of soft computing encompasses fuzzy logic and other meta-heuristics such as genetic algorithms, neural networks, simulated annealing, particle swarm optimization, ant colony systems and parts of learning theory as shown in Figure 2.1 (Chen 2002). It is expected that soft computing techniques have received increasing attention in recent years for their interesting characteristics and their success in solving problems in a number of fields.

The Components of soft computing is shown in figure 1.

Soft computing techniques have been applied to text document clustering as an optimization problem. An innovative field has developed many hybrid ways of optimization techniques such as PSO and ACO with FCM and K-means [6, 17, 18, 41] to further improve efficiency of document clustering.

In traditional clustering techniques (k-means and their variants, FCM, etc.) for clustering, some parameters must be known in advance like number of clusters. By using optimization techniques clustering quality can be improved.

The most popularly used optimization techniques to document clustering are listed below:

### 2.1 Genetic Algorithm (GA)

The Genetic Algorithm (GA) [12] is a very popular evolutionary algorithm that was first pioneered by John Holland in 1970s. The basic idea of GAs is designed to make artificial systems software that retains the robustness of natural biological evolution system. Genetic algorithms belong to search techniques that mimic the principle of natural selection. The performance of GA is influenced by the following three operators:

1. **crossover**
2. **mutation**
3. **selection**

The selection operator is based on a fitness function; parents are selected for mating (i.e. recombination or crossover) according to their fitness. Based on the selection operator, the better the chromosomes would be included in the next populations and the others would be eliminated. The popularly used methods that can be used to select the best chromosomes are roulette wheel selection, tournament selection, Boltzman selection, Steady state selection and rank selection. The crossover is applied and it selects genes from parent chromosomes in order to create the new population. The crossover point is selected with probability Pc. After a crossover is performed, mutation takes place. A mutation can be applied by randomly modifying bits. The proportion of mutation is probability-based such as Pm. The working principle of GA is given below:

1. Generate initial population;
2. Evaluate population;
3. Repeat until stopping criterion is met

{
    a. Select parents for reproduction using selection operator;
    b. Perform crossover and mutation using genetic operator;
    c. Evaluate population fitness function;
}

### 2.2 Bees Algorithm (BA)

Another population-based search algorithm is the bees algorithm [35]. The bees algorithm [44] is an optimization algorithm first developed in 2005 and it is based on the natural food foraging activities of honey bees. As a first step, an initial population of solutions are created and then evaluated based on a fitness function. Then based on this evaluation, the highest finesses values are selected for neighbourhood search, assigning more bees to search bear to the best solutions of the highest fitness's. Then for each patch select the fittest solution to build the next new population. The remaining bees in the population are allocated randomly around the search space scouting for new possible solutions. These steps are repeated until a stopping criterion is met.
The Algorithm steps are given below:

1. Generate initial population solutions randomly (n).
2. Evaluate fitness of the population.
3. Continue below steps until stopping criterion is met
    {
    a. Choose highest fitnesses (m) and allow them for neighborhood search.
    b. Recruit more bees to search near to the highest fitness's and evaluate fitnesses.
    c. Select the fittest bee from each patch.
    d. Remaining bees are assigned for random search and do the evaluation.
    }

The proper values for the parameters n,m are set.

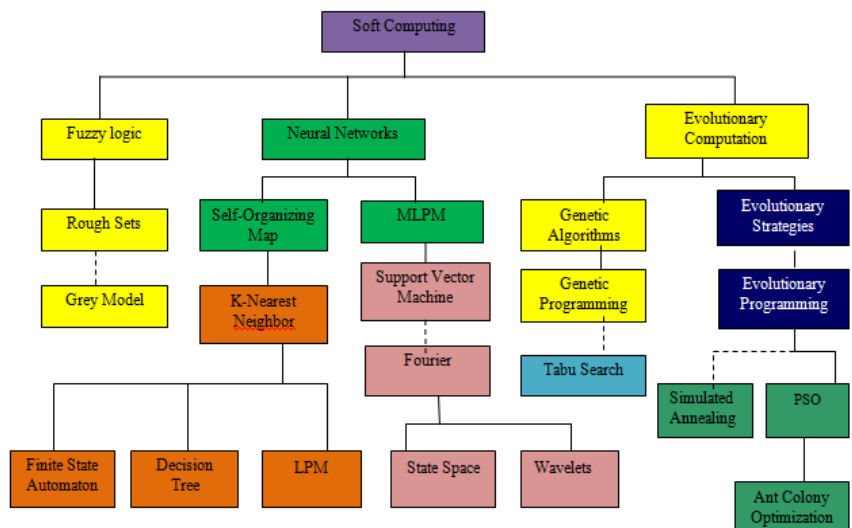

Figure 1. The hybridization of soft computing

## 2.3 Particle Swarm Optimization (PSO)

Particle Swarm Optimization (PSO) [14] [15] is also an evolutionary computation technique. It is enthused by study of flocks of birds' behaviours. A PSO algorithm starts by having a population of candidate solutions. The population is called a swarm and each solution is called a particle. Initially particles are assigned a random initial position and an initial velocity. The position of each particle has a value obtained by the objective function. Particles are moved around in the search-space. While moving in the search space, each particle remembers the best solution it has achieved so far, called as the individual best fitness (pbest), and the position of the best solution, called as the individual best fitness position. Also, it maintains the value that can be obtained so far by any particle in the population, referred to as the global best fitness (gbest) and its position is referred to as the global best fitness position. During this operation when improved positions are being found then these positions will come to guide the movements of the swarm. In each iteration, particles are updated according to two "best" values. They are called pbest and gbest. After calculating the two best values, the particle updates its velocity and positions by using the following equations:

$$vel[\ ] = vel[\ ] + l_1 * rand() * (pbest[\ ] - current[\ ]) + l_2 * rand() * (gbest[\ ] - current[\ ])$$
$$current[\ ] = current[\ ] + vel[\ ]$$

where

    vel[ ]     : the particle velocity
    current [ ] : the current particle (solution)
    pbest[ ]   : personal best value
    gbest[ ]   : global best value
    rand ()    : random number between (0,1)
    $l_1, l_2$     : learning factors.

The following steps are done in PSO algorithm:

1. Initialize each particle in the population with random positions and velocities.
2. Repeat the following steps until stopping criterion is met.

   i. for each particle

```
{
    Calculate the fitness function value;
    Compare the fitness value: If it is superior to the best fitness value pbest,
    then current value is assigned pbest value;
}
```

   ii. Best fitness value particles among all the particles are selected and assign it as gbest;
   ii. for each particle

```
{
    Calculate particle velocity;
    Change the position of the particle;
}
```

## 2.4 Ant Colony Optimization (ACO)

Natural ant behaviour was first modelled by using Ant-based clustering algorithms. Ant Colony Optimization was initially inspired by Deneubourg in 1990 in aiming to search for an optimal path in a graph, based on the behaviour of ants .

In nature, ants find the shortest path between their hole and a source of food. In order to show their paths, Ants use a substance called Pheromone. The ant colony optimization algorithm [21] presents a random search method. Initially, a two-dimensional m*m space is considered. The size of this space is selected larger enough to hold the clustered elements. According to the cumber of data objects to be clustered, the number of ants and the value m will be determined. For example if n is the number of data elements to be clustered, then the number of ants will be m=4n, n/3. The repetitive process includes a stack of elements which are created based on the idea that the ant collects the similar adjacent elements. Then larger stacks are created by combining smaller stacks. In the end, clusters are the final obtained stacks.

In the following sections document representation model for clustering, weighting scheme, evaluation criteria and dimension reductions are discussed. Next, we also put forward optimization approaches related to document clustering as a segment in the survey.

## 3. DOCUMENT CLUSTERING

Clustering of documents is used to group documents into relevant topics. The major difficulty in document clustering is its high dimension. It requires efficient algorithms which can solve this high dimensional clustering. A document clustering is a major topic in information retrieval area .Example includes search engines. The basic steps used in document clustering process are shown in figure 2.

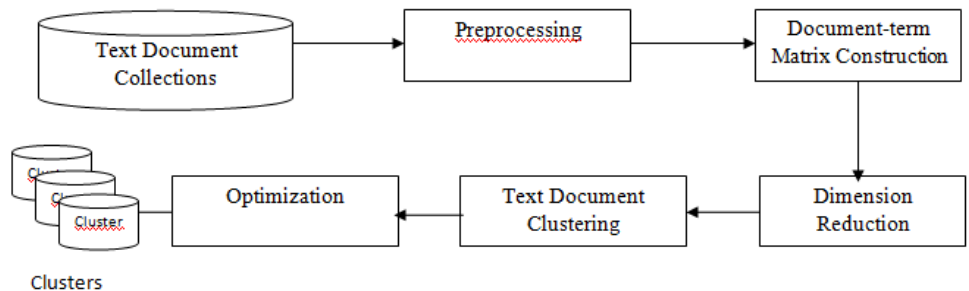

Figure 2.Flow diagram for representing basic Steps in text clustering

### 3.1 Peprocessing

The text document preprocessing basically consists of a process to strip all formatting from the article, including capitalization, punctuation, and extraneous markup (like the dateline,tags). Then the stop words are removed. Stop words term (i.e., pronouns, prepositions, conjunctions etc) are the words that don't carry semantic meaning. Stop words can be eliminated using a list of stop words. Stop words elimination using a list of stop word list will greatly reduce the amount of noise in text collection, as well as make the computation easier. The benefit of removing stop words leaves us with condensed version of the documents containing content words only.

The next process is to stem a word. Stemming is the process for reducing some derived words into their root form. For English documents, a popularly known algorithm called the Porter stemmer [7] is used. The performance of text clustering can be improved by using Porter stemmer.

### 3.2 Text Document Encoding

The next process is to encode the text document. In general, documents are transformed into document term matrix (DTM) which is a mathematical matrix whose dimensions are the terms and rows are documents. A simple DTM is the vector space model [1] model which is widely used in IR and text mining [23], to represent the text documents. It is used in indexing, information retrieval and relevancy rankings and can be successfully used in evaluation of search results from web search engines.

Let $\mathbf{D} = (D_1, D_2, \ldots, D_N)$ be a collection of documents and $\mathbf{T}=(T_1, T_2, \ldots, T_M)$ be the collection of terms of the document collection $\mathbf{D}$, where N is the total number of documents in the collection and M is the number of distinct terms. In this model each document $D_i$ is represented by a point in an m dimensional vector space, $D_i = (w_{i1}, w_{i2}, \ldots, w_{im})$, $i = 1, \ldots, N$, where the dimension is the total number of unique terms in the document collection. Many schemes have been proposed for measuring $w_{ij}$ values, also known as (term) weights. One of the more advanced term weighting schemes is the tf-idf (term frequency-inverse document frequency) [23]. The tf-idf scheme aims at balancing the local and the global weighting of the term in the document and it is calculated by

$$w_{ij} = tf_{ij} \times \log\left(\frac{n}{df_j}\right) \qquad (1)$$

where $tf_{ij}$ is the frequency of term i in document j, and $df_{ij}$ denotes the number of documents in which term j appears. The component log $(n/df_{ij})$, which is often called the idf factor, defines the global weight of the term j.

### 3.3 Dimension reduction techniques

Dimension reduction can be divided into feature selection and feature extraction. Feature selection is the process of selecting smaller subsets (features) from larger set of inputs and Feature extraction transforms the high dimensional data space to a space of low dimension. The goals of dimension reduction methods are to allow fewer dimensions for broader comparisons of the concepts contained in a text collection.

In this paper, one dimension reduction technique is discussed.

### 3.3.1 Latent Semantic Indexing (LSI)

A popular text retrieval technique is Latent Semantic Indexing, which uses Singular value decomposition (SVD) to identify patterns in the relationships between the terms and concepts contained in a collection of text. The singular value decomposition reduces the dimensions by selecting dimensions with highest singular values. For text processing LSI can be effectively used because it preserves the polysemy and synonymy in the text. A key feature of LSI is its ability to retain latent structure in word and hence it improves the clustering efficiency.

Once a term-document matrix X (M×N) is constructed, assuming there are m distinct terms and n documents. The Singular Value Decomposition computes the term and document vectors by transforming TDM matrix X into three matrices P, S and D, which is given by

$$X = PSQ^T \qquad (2)$$

where

    P : left singular vector matrix
    Q : right singular vector matrix
    S : diagonal matrix of singular values.

LSI approximates X with a rank *k* matrix.

$$X_k = P_k S_k Q^T_k \qquad (3)$$

where $P_k$ is defined to be the first *k* columns of the matrix P and $Q^T_k$ is included the first *k* rows of matrix $Q^T$. $S_k = \text{diag}(s_1, \ldots, s_k)$ is the first *k* largest singular values.

When LSI is used for text document clustering, a document $D_i$ is represented by [11]

$$D_i = D^T_i P_k \qquad (4)$$

Then the text corpus can be organized by another representation of document-term matrix D(N×M) and the corpus matrix is organized by

$$C = DP_k \qquad (5)$$

## 3.4 Similarity Measurement

The similarity (or dissimilarity) between the objects is typically computed based on the distance between document pairs. The most popular distance measure as stated by [3] are:

Table 1. Formulas for Similarity Measurement

| Name | Formula |
|---|---|
| Euclidean Distance | $\sqrt{(x_{i1}-x_{j1})^2 + (x_{i2}-x_{j2})^2 + \ldots + (x_{in}-x_{jn})^2}$ where $i=(x_{i1},x_{i2},\ldots,x_{in})$ and $j=(x_{j1},x_{j2},\ldots,x_{jn})$ |
| Manhattan distance | $\lvert(x_{i1}-x_{j1})\rvert + \lvert(x_{i2}-x_{j2})\rvert + \cdots = \lvert(x_{in}-x_{jn})\rvert$ |
| Minkowski distance | $\left(\lvert(x_{i1}-x_{j1})^p\rvert + \lvert(x_{i2}-x_{j2})^p\rvert + \cdots + \lvert(x_{in}-x_{jn})^p\rvert\right)^{1/p}$ where p is a positive integer |
| Jaccard coefficient | $J(A,B) = (A \cap B)/(A \cup B)$ where A and B are documents |
| Cosine Similarity | $S(x,y) = \dfrac{x^t \cdot y}{\lVert x \rVert \lVert y \rVert}$, where $x^t$ is a transposition of vector x, $\lVert x \rVert$ is the Euclidean norm of vector x, $\lVert y \rVert$ is the Euclidean norm of vector y |

## 3.5 Evaluation of Text Clustering

The quality of text clustering can be measured by using the popularly used external indexes [22]: F-measure, Purity and Entropy. These measures are called external quality measures because the results of clustering techniques are compared with known classes (i.e) it requires that the documents be given class labels in advance. A widely used quality measures for the purpose of text document clustering [22] are F-measure, Purity and Entropy which can be defined as follows:

### 3.5.1. F-measure

The F-measure combines the precision and recall values used in information retrieval. The *precision P(i,j)* and *recall R(i,j)* of each cluster j for each class i are calculated as

$$P(i,j) = \frac{\beta_{ij}}{\beta_j} \tag{6}$$

$$P(i,j) = \frac{\beta_{ij}}{\beta_i} \tag{7}$$

where

$\beta_i$ : is the number of members of class i
$\beta_j$ : is the number of members of cluster j
$\beta_{ij}$: is the number of members of class i in cluster j

The corresponding *F-measure F(i,j)* is given by the following equation:

$$F(i,j) = \frac{2 * P(i,j) * R(i,j)}{P(i,j) + R(i,j)} \tag{8}$$

Then the *F-measure* of a class i can be defined as

$$F = \sum_i \frac{\beta_i}{n} \max_j (F(i,j)) \tag{9}$$

where n is the total number of documents in the collection. In general, the larger the F-measure gives the better clustering result.

### 3.5.2. Purity

The purity measure of a cluster represents the percentage of correctly clustered documents and thus the purity of a cluster j is defined as

$$Purity(j) = \frac{1}{\beta_j} \max_i (\beta_{ij}) \tag{10}$$

The overall purity of a clustering is a weighted sum of the cluster purities and is defined as

$$Purity = \sum_n \frac{\beta_j}{n} Purity(j) \tag{11}$$

In general, the better clustering result is given by the larger the purity value.

### 3.5.3. Entropy

The Entropy of a cluster can be defined as the degree to which each cluster consists of objects of a single class. The entropy of a cluster j is calculated using the standard formula,

$$e_j = -\sum_{i=1}^{L} p_{ij} \log p_{ij} \qquad (12)$$

where
    L : Number of classes
    $p_{ij}$: Probability that a member of cluster j belongs to class i.

The total entropy of the overall clustering result is defined to be the weighted sum of the individual entropy value of each cluster. The total entropy e is defined as

$$e = \sum_{j=1}^{k} \frac{\beta_j}{n} e_j \qquad (13)$$

where
    k : Number of clusters
    n : Total number of documents in the corpus.

In general, the better clustering result is given by the smaller entropy value.

### 3.6 Datasets

For evaluating the effectiveness of the document clustering algorithms, several text collections are available. These collections are useful for research in information retrieval, natural language processing, computational linguistics and other corpus-based research. The following real text datasets have been selected for clustering purpose. The datasets are:

*Reuters-21578:* Reuters-21578 test collection contains 21578 text documents. The documents in the Reuters-21578 collection are originally taken from Reuters newswire in 1987. The Reuters-21578 contains 22 files. Each of the first 21 files (reut2-000.sgm through reut2-020.sgm) contains 1000 documents, while the last (reut2-021.sgm) contains 578 documents. The documents are broadly divided into five broad categories (Exchanges, People, Topics, Organizations and Places). These categories are further divided into subcategories. The Reuters-21578 test collection is available at [9].

*20NewsGroup:* 20 Newsgroups data set contained 20,000 newsgroup articles from 20 newsgroups on a variety of topics. The dataset is available in http://www.cs.cmu.edu/afs/cs/project/theo-20/www/data/news20.html. It is also available in the UCI machine learning dataset repository available at http://mlg.ucd.ie/datasets/20ng.html. This dataset was assembled by Ken Lang.

*Hamshahri* : Hamshahri collection was used for evaluation of Persian information retrieval systems. Hamshahri collection consists of about 160,000 text documents and 100 queries in Persian and English languages. There are totally 50350 query document pairs upon which relevance judgements are made. The judgement result is binary judges, either "1" (relevant) or "0" (irrelevant).

## 4. RELATED WORKS

Document clustering had been widely studied in computer science literature. Several soft computing techniques have been used for text document clustering and some of these are discussed here.

Wei Song, Soon Cheol Park [11] proposed a variable string length genetic algorithm for automatically evolving the proper number of clusters as well as providing near optimal data set clustering. GA was used in conjunction with the reduced latent semantic structure to improve clustering efficiency and accuracy. The objective function used was Davis-Bouldin index. It produces much lower cost of computational time due to the reduction of dimensions.

K. Premalatha, A.M. Natarajan [17] presented a new hybrid model of clustering based on GA and PSO which were used to solve document clustering problem. The use of Genetic Algorithm in this model is to attain an optimal solution. Due to the simplicity of PSO and efficiency in navigating large search spaces for optimal solution, it is combined with GA. This hybrid model avoided the premature convergence and improved the diversity.

Eisa Hasanzadeh [27], [32] developed a text clustering algorithm based on PSO and applied latent semantic indexing (PSO+LSI) for reducing dimension. Latent Semantic Indexing (LSI) was used to reduce the high dimension of textual data. Because of the main problem of text clustering algorithm is very high dimension; it is avoided by using LSI. PSO family of bio-inspired algorithms had successfully been merged with LSI. [27] used an adaptive inertia weight (AIW) that does proper exploration and exploitation in search space. This model produced better clustering results over PSO+Kmeans using vector space model. The proposed work PSO+LSI are faster than PSO+Kmeans algorithms using the vector space model for all numbers of dimensions.

Stuti Karol , Veenu Mangat [43] introduced hybrid PSO based algorithm. The two partitioning clustering algorithms Fuzzy C-Means (FCM) and K- Means each hybridized with Particle Swarm Optimization (KPSO and FCPSO). The performance of hybrid algorithms provided better document clusters against traditional partitioning clustering techniques (K-Means and Fuzzy C Means) without hybridization. It is concluded that FCPSO deals better with overlapping nature of dataset than KPSO as it deals well with the overlapping nature of documents.

Nihal M. AbdelHamid, M. B. Abdel Halim, M. Waleed Fakhr[36]  introduced the Bees Algorithm in optimizing the document clustering problem. The Bees algorithm avoids local minima convergence by performing global and local search simultaneously. This proposed algorithm has been tested on a data set containing 818 documents and the results have revealed that the algorithm achieved its robustness. This model was compared with the Genetic Algorithm and K-means and it was concluded that Bees algorithm outperforms GA by 15% and the K-means by 50%. And also the results revealed that the Bees Algorithm takes more time than the Genetic Algorithm by 20% and the K-means by 55%.

Kayvan Azaryuon,Babak Fakhar [45] proposed an upgraded the standard ant's clustering algorithm by changing the Ants' movement completely random for clustering. This model provided increasing quality and minimizing run time when compared with the standard ant clustering algorithm and the K-means algorithm. His proposed algorithm has been implemented to cluster documents in the Reuters-21578. The Results have shown that the proposed algorithm presents a better average performance than the standard ants clustering algorithm, and the K-means algorithm.

Table 2. Comparison of various Optimization based document clustering methods

| Paper referred | Dimension Reduction applied | Dataset used | Evaluation parameters | Fitness function used |
|---|---|---|---|---|
| Wei Song, Soon Cheol Park [11] | Yes | Reuters-21578 | F-measure | Davis-Bouldin Index |
| K. Premalatha, A.M. Natarajan[17] | No | Silvia et al 2003 | Fitness value | $h_f = \frac{1}{N_c} \sum_{i=1}^{N_c} \frac{S_i}{P_i}$ |
| Eisa Hasanzadeh, Morteza Poyan rad and Hamid Alinejad Rokny[27] [32] | Yes | Reuters-21578 Hamshahri | F-measure | ADDC (Average Distance Documents to the cluster Centroid) |
| Stuti Karol, Veenu Mangat[43] | No | Reuters-21578 | F-measure Entropy | ADDC (Average Distance Documents to the cluster Centroid) |
| Nihal M. AbdelHamid, M. B. Abdel Halim, M. Waleed Fakhr [36] | No | Web Articles: www.cnn.com www.bbc.com news.google.com www.2knowmyself.comm | Fitness value | Cosine Distance |
| Kayvan Azaryuon, Babak Fakhar[45] | No | Reuters-21578 | F-measure | --- |

## 5. CONCLUSION

This paper has presented a survey on the research work done on text document clustering based on optimization techniques. This survey starts with a brief introduction about clustering in data mining, soft computing and explored various research papers related to text document clustering. More research works have to be carried out based on semantic to make the quality of text document clustering.

International Journal on Computational Sciences & Applications (IJCSA) Vol.3, No.6, December 2013

**Authors**

**R.Jensi** received the B.E (CSE) and M.E (CSE) in 2003 and 2010, respectively. Her research areas include data mining, text mining especially text clustering, natural language processing and semantic analysis. She is currently pursuing Ph.D(CSE) in Manomanium Sundaranar University , Tirunelveli. She is a member of ISTE.

**Dr.G.Wiselin Jiji** received the B.E (CSE) and M.E (CSE) degree in 1994 and 1998, respectively. She received her doctorate degree from Anna University in 2008. Her research interests include classification, segmentation, medical image processing. She is a member of ISTE.